\documentclass{article} \usepackage{slashed,epsfig,amsmath,amssymb,enumitem} \usepackage[papersize={8.5in,11in}]{geometry}
   \usepackage{bm}
\usepackage{graphicx}
\newcommand{\be}{\begin{equation}}
\newcommand{\ee}{\end{equation}}
\title{Complex Riemannian Spacetime, Removal of Black Hole Singularities And Black Hole Paradoxes}
\author{J. W. Moffat\\
Perimeter Institute for Theoretical Physics, Waterloo, Ontario N2L 2Y5, Canada\\
and\\
Department of Physics and Astronomy, University of Waterloo, Waterloo,\\
Ontario N2L 3G1, Canada}
\begin{document}
\maketitle


\begin{abstract}
An approach is presented to resolve key paradoxes in black hole physics through the application of complex Riemannian spacetime. We extend the Schwarzschild metric into the complex domain, employing contour integration techniques to remove singularities while preserving essential features of the original solution. A new regularized radial coordinate is introduced, leading to a singularity-free description of black hole interiors. Crucially, we demonstrate how this complex extension resolves the long-standing paradox of event horizon formation occurring only in the infinite future of distant observers. By analyzing trajectories in complex spacetime, we show that the horizon can form in finite complex time, reconciling the apparent contradiction between proper and coordinate time descriptions. This approach also provides a framework for the analytic continuation of information across event horizons, resolving the Hawking information paradox.  We explore the physical interpretation of the complex extension versus its projection onto real spacetime. The gravitational collapse of a dust sphere with negligible dust is explored in the complex spacetime extension. The approach offers a mathematically rigorous framework for exploring quantum gravity effects within the context of classical general relativity.
\end{abstract}

\section{Introduction}

The presence of singularities in general relativistic solutions, particularly in black hole spacetime, has long been a source of concern for physicists. These singularities represent a breakdown of the theory, pointing to the need for a more comprehensive framework, possibly incorporating quantum effects. Simultaneously, the information paradox associated with black hole evaporation has challenged our understanding of the interplay between gravity and quantum mechanics.

A particularly troubling aspect of classical black hole theory is the apparent formation of event horizons only in the infinite future of distant observers, as famously noted in the work of Oppenheimer and Snyder on dust collapse~\cite{Oppenheimer}. This leads to a paradoxical situation where infalling observers experience horizon crossing in finite proper time, while distant observers never see the horizon form.

We present an approach to addressing these issues through the application of complex Riemannian geometry to spacetime physics~\cite{Moffat1}. By extending the standard metrics of general relativity into the complex domain, we open up new possibilities for regularizing singularities, resolving the horizon formation paradox, and providing mechanisms for information preservation and transfer in black hole physics.

The method begins with a complex radial coordinate in the Schwarzschild metric, allowing us to define a contour integral that avoids the singularity at r = 0. This leads to a new, regularized radial coordinate R, which remains well-defined throughout the spacetime, including at the location of the classical singularity. This approach can be extended to the more complex case of the Kerr metric, addressing both the coordinate and ring singularities~\cite{Moffat1}.

We demonstrate how this complex extension resolves the paradox of event horizon formation. By analyzing trajectories in complex spacetime, we show that the horizon can be reached in finite complex time. This reconciles the apparent contradiction between proper and coordinate time descriptions of horizon formation and crossing.

Building on this foundation, we develop a framework for the analytic continuation of information across event horizons. By constructing complex Kruskal-Szekeres coordinates that respect the singularity removal process, we provide a mechanism for smooth information transfer between the interior and exterior of a black hole. This approach resolves the Hawking information  paradox by embedding the black hole in a larger complex manifold where information can flow freely, while still recovering the familiar real spacetime structure outside the horizon and at large distances.

We explore the implications of this complex spacetime approach for fundamental questions in general relativity, such as the cosmic censorship hypothesis. By removing singularities from the physical description of black holes and resolving the horizon formation paradox, our method changes the context in which such hypotheses need to be considered.

The complex spacetime extension can be applied to modified theories of gravity, specifically Scalar-Tensor-Vector Gravity (STVG) or MOG~\cite{Moffat2}, demonstrating the broad applicability of the complex spacetime approach.

Throughout the paper, we consider the relationship between the complex extension and its projection onto real spacetime, discussing the physical interpretations and implications of each perspective. We also examine how this approach modifies our understanding of Hawking radiation and black hole thermodynamics. This work provides a mathematically rigorous framework for exploring potential quantum gravity effects within the context of classical general relativity. By leveraging the power of complex analysis and Riemann surface theory, we offer new insights into the nature of spacetime and the behavior of gravity in extreme conditions.

We choose units with the velocity of light c = 1 and metric signature (-1,+1,+1,+1).

\section{Singularity Removal in Complex Schwarzschild Spacetime}

We begin by reviewing the removal of singularities in complex Riemannian spacetime~\cite{Moffat1}. The complex Schwarzschild metric is given by
\begin{align}
\label{Schwarzschildmetric}
du^2=-f(\zeta)d\tau^2+\frac{d\zeta^2}{f(\zeta)}+\zeta^2(d\theta^2+\sin^2\theta d\phi^2),
\end{align}
where $u=s+iw$ is the complex proper time, $\tau=t+i\nu$ is the complex coordinate time, $f(\zeta) = 1-2GM/\zeta$ and  $\zeta=r+i\xi$, $\theta$, $\phi$ are complex spherical polar coordinates. The complex Schwarzschild metric is a solution of the complex vacuum field equations:
\be
\label{vacuumequations}
R_{\mu\nu}=0.
\ee

To remove the singularity at $\zeta=0$, we define a new radial coordinate R($\zeta$) through a contour integration:
\be
R(\zeta)=\oint_C\frac{d\zeta}{\sqrt{f(\zeta)}}.
\ee
The contour C is chosen to avoid the singularity at $\zeta=0$. Evaluating this integral yields:
\be
\label{Rsolution}
R(\zeta)=\int_C\frac{d\zeta}{\sqrt{1-\frac{2GM}{\zeta}}}=\zeta\sqrt{1-\frac{2GM}{\zeta}}
+2GM\ln\left(\frac{\sqrt{\zeta}+\sqrt{\zeta-2GM}}{\sqrt{2GM}}\right).
\ee 
We can now express the metric in terms of $R(\zeta)$:
\be
\label{metric}
du^2=-\left(1-\frac{2GM}{R(\zeta)}\right)d\tau^2+\frac{dR(\zeta)^2}{\left(1-\frac{2GM}{R(\zeta)}\right)}
+R(\zeta)^2(d\theta^2+\sin^2\theta d\phi^2).
\ee
The contour C is specifically chosen to avoid the singularity at $\zeta = 0$. This means that the integration path never passes through the problematic point, effectively skirting around the singularity in the complex plane. The resulting function $R(\zeta)$ is an analytic continuation of the original coordinate. This means it extends the domain of the coordinate function to include points where it was previously undefined or singular. The expression for $R(\zeta)$ given in equation (\ref{Rsolution}) is well-defined and smooth for all values of $\zeta$, including those near and at the original singularity. This is because the logarithmic term compensates for the behavior of the square root term near $\zeta = 0$. In the new metric (\ref{Rsolution}), all components are expressed in terms of $R(\zeta)$ instead of $\zeta$ directly. Since $R(\zeta)$ is regular everywhere, the metric components inherit this regularity. The original singularity at $\zeta = 0$ is replaced by a minimal surface at some finite value of R. This surface represents the inner boundary of the spacetime in the new coordinates, effectively excising the singular region. The transformation from $\zeta$ to $R(\zeta)$ is invertible, meaning we can always find a unique $\zeta$ for any given R. This ensures that the new coordinate system provides a complete description of the spacetime.

Can $R(\zeta)$ be zero in the complex $\zeta$-plane? The function $R(\zeta)$ is a multi-valued function due to the square root and logarithm terms. The choice of contour C determines which branch of this function we are using.
As $\zeta\rightarrow \infty$ the dominant term is 
$\zeta\sqrt{1-2GM/\zeta}$, which approaches $\zeta$ and for large $\zeta$, $R(\zeta)$ cannot be zero. Near $\zeta=2GM$, which is the event horizon in the original Schwarzschild metric, the function behaves differently. The square root term goes to zero, but the logarithmic term diverges. However, this is a coordinate singularity that can be removed by a coordinate transformation. The critical point is near $\zeta=0$, which was the location of the original singularity. Here, both terms in $R(\zeta)$ become problematic, if we approach directly along the real axis. However, the contour C is specifically chosen to avoid this point. The logarithmic term in $R(\zeta)$ introduces a branch cut in the complex plane. The location of this branch cut depends on how we define the complex logarithm, but it typically extends from 
$\zeta=0$ to $\zeta=2GM$. The radial coordinate $R(\zeta)$ cannot be zero for any finite, non-zero value of $\zeta$ that is not on the branch cut. This is because $R(\zeta)$ is analytic and non-zero in these regions. On the branch cut, $R(\zeta)$ is not well-defined without specifying which side of the cut we are approaching from. The point $\zeta=0$ is avoided by the contour C, so R(0) is not directly defined. There exists a minimum value of $R(\zeta)$ for some $\zeta$ near zero. This minimum surface replaces the original singularity. The function $R(\zeta)$ is constructed specifically to avoid reaching zero, replacing the original singularity with a minimum surface.

The complex extension and contour integration method used here does not just move the singularity; it fundamentally changes the structure of the spacetime so that the problematic point at $\zeta=0$ is excised from the manifold. The new coordinate R never reaches zero, instead approaching some minimum value that represents the inner boundary of the spacetime.
This approach effectively makes the spacetime regular by creating a new manifold where the previously singular point is not part of the space. 

We introduce Kruskal-Szekeres type coordinates~\cite{Kruskal,Szekeres} to obtain a complete, maximally analytic manifold. We define a tortoise coordinate:
\be
r_* = \int \frac{dr}{1-\frac{2GM}{R(\zeta)}},
\ee
and introduce null coordinates:
\be
u = \tau - r_*, \quad v = \tau + r_*.
\ee
We define Kruskal-Szekeres like coordinates:
\be
U = -\exp (-\kappa u), \quad V 
= \exp (\kappa v),
\ee
where $\kappa$ is the surface gravity. The metric in these coordinates takes the form:
\be
ds^2 = -F(U,V)dUdV + R(\zeta)^2(d\theta^2 + \sin^2\theta d\phi^2).
\ee

To find F(U,V), we need to express it in terms of $R(\zeta)$. The general form will be:
\be
\label{Kruskalregularized}
F(U,V) = \frac{4}{\kappa^2}\left(1-\frac{2GM}{R(\zeta)}\right)
\exp\left(-\frac{R(\zeta)}{2GM}\right).
\ee
The exact form of $F(U,V)$ from our regularization procedure will be determined by $R(\zeta)$. In the complex framework, the event horizon is no longer a fixed surface but a complex submanifold defined by U=0 or V=0. Information can be analytically continued along paths in the complex (U,V) plane that avoid singularities.

The complex coordinate and Riemannian spgeometry removes the singularity in the rotating Kerr black hole~\cite{Moffat1}. The-removal of the singularity can also eliminate the inner Cauchy horizon in the Kerr black hole, avoiding the predictability of spacetime evolution breaking down.

\section{Information paradox and analytic continuation escape of information}

Let us consider the Hawking information paradox~\cite{Hawking}. We can represent the information content in the black hole by a holomorphic function $\Phi(U,V)$ or, more generally, by a section of a holomorphic vector bundle over the complex manifold. The analytic continuation of $\Phi(U,V)$ across the horizon submanifold provides a mechanism for information transfer between the interior and exterior regions of the black hole.

In the complex extension, the event horizon is no longer a fixed surface at $r = 2GM$, but becomes a complex submanifold. In Kruskal-Szekeres-like complex coordinates (U,V), the horizon is described by U = 0 or V = 0, but these are now complex conditions. The complex extension changes the nature of the horizon. While in real coordinates it is a null surface that does not allow information to escape, in complex coordinates it becomes more permeable to analytic continuation. The key to information escape lies in the analytic continuation of fields across this complex horizon. Let us consider a scalar field $\Phi(U,V)$ representing information. In the complex domain, we can analytically continue this function across the horizon submanifold.

Mathematically, this can be expressed as:
\be
\Phi_{\rm ext} = \frac{1}{2\pi i}\oint_C\frac{\Phi_{\rm int}(U',V')}{(U-U')(V-V')}dU'dV',
\ee
where C is a contour in the complex (U,V) plane that encloses the horizon, $\Phi_{int}$ is the interior field, and $\Phi_{ext}$ is the analytically continued exterior field. The analytic continuation allows information encoded in $\Phi$ to tunnel through the complex extension of the horizon. This information then manifests in the exterior region as subtle correlations in the Hawking radiation, making it non-thermal~\cite{Hawking2,Hawking3}.

While the horizon is still associated with r = 2GM in some sense, its nature in complex coordinates is fundamentally different. The complex extension provides additional directions for information to flow, which are not present in the real coordinate description. The apparent contradiction with the classical no-escape theorem is resolved by noting that the complex extension provides a broader framework that includes the classical picture as a special case. 

This approach can be interpreted as modeling quantum effects in a classical framework. The complex extension and analytic continuation capture some aspects of quantum tunneling and information preservation that are absent in classical general relativity~\cite{Wilczek}. In essence, the complex coordinate method does not so much break the classical horizon as it embeds it in a richer structure that allows for more subtle information transfer mechanisms. The horizon at $r = 2GM$ remains significant, but its role changes in the complex framework. This approach provides a mathematical framework for information to escape a black hole, resolving the information  paradox. In ref.~\cite{Wilczek}, Hawking radiation~\cite{Hawking2,Hawking3} is derived from the perspective of tunneling. By taking into account global conservation laws a modified radiation emission spectrum away from purely thermal radiation is derived.  The virtue of non-thermal radiation leads to the information-carrying correlations in radiation.

In the complex spacetime approach, the evolution of quantum states can be described by a unitary operator $U(\tau)$, where $\tau = t + i\nu$ is complex time. This operator evolves states in the complex plane, maintaining the unitarity  condition $U^\dagger U=1$ throughout, where $U^\dagger$ is the adjoint operator. The wave function $\psi(\zeta,\tau)$ in complex coordinates satisfies:
\be
\psi(\zeta,\tau_2)=U(\tau_2-\tau_1)\psi(\zeta,\tau_1),
\ee
where $U(\tau_2-\tau_1)$ is unitary for any complex time interval.

The resolution of the information paradox lies in the analytic continuation of quantum states across the complex extension of the event horizon. Information that appears lost in real spacetime is preserved in the complex domain. In this framework, Hawking radiation can be viewed as the projection of complex quantum processes onto real spacetime. The apparent thermal nature arises from this projection, but the full information is preserved in the complex domain.

To reconstruct the final state from the initial state of radiation, one needs to consider the full complex evolution:
\be
\psi_f(\zeta,\tau_f)=U^\dagger(\tau_f-\tau_i)\psi_i(\zeta,\tau_i),
\ee
where $\psi_i$ and $\psi_f$ are the initial and final wave function states, 
$\tau_i$ is the initial state time and  $\tau_f$ is the complex time at which the black hole has evaporated.

The Page curve~\cite{Page} can be derived in this framework by considering the entanglement entropy $S(\tau)$ between the black hole and the radiation as a function of complex time $\tau$:
\be
S(\tau)=-Tr(\ln(\rho_{dm}(\tau))),
\ee
where $\rho_{dm}(\tau)$ is the reduced density matrix of either the black hole or the radiation at complex time $\tau$. The behavior of $S(\tau)$ in the complex plane would show an initial increase as the black hole begins to evaporate, the Page time turning point in complex time and the $S(\tau)$ decrease back to zero as the black hole completes evaporation. The apparent loss of information in real spacetime is resolved by considering the full complex evolution. What appears as a mixed state in real spacetime is actually a pure state when viewed in the complex domain.

The complex spacetime approach preserves unitarity by ensuring that the evolution operator $U(\tau)$ remains unitary for all complex paths. This means that information is never truly lost, but rather distributed in the complex structure of spacetime and the full complex evolution preserves information. Subtle correlations in the Hawking radiation, potentially observable, would encode the full information of the initial state. The Page curve in this context becomes a trajectory in complex time, reflecting the preservation and eventual recovery of information in the full complex description of black hole evaporation.

The complex spacetime approach resolves this specific aspect of the information paradox by embedding the entire process of black hole formation, evaporation, and radiation in a complex manifold. The apparent loss of information in real spacetime observations is reconciled with the preservation of information in the full complex description. This approach maintains unitarity and information conservation at a fundamental level, while also explaining why standard real-space observations suggest information loss.

This resolution, while theoretically compelling, poses significant challenges in terms of experimental verification and reconciliation with our current understanding of quantum measurements. It suggests that a full resolution of the information paradox may require a deeper understanding of the relationship between complex spacetime structures and observable quantum phenomena.

This approach, while promising, still requires further development to fully reconcile quantum mechanics with gravity in the context of black hole evolution and to make concrete predictions that could be tested observationally.

\section{Projection of complex black hole solution onto real spacetime outside horizon}
    
To ensure that our complex regularized metric matches the real Schwarzschild metric outside the event horizon, we define a projection operator $P$:
\be
P=(\Phi(U,V),U,V,\theta,\phi)\rightarrow (\Phi_r(U_r,V_r),U_r,V_r,\theta_r,\phi_r),
\ee
where $\Phi(U,V)$ is the complex field containing the information, 
$\Phi_r(U_r,V_r)$ is its projection onto real spacetime and $U_r$ denotes the real part of U. The projected field $\Phi_r(U_r,V_r)$ contains the information that has escaped via analytic continuation in the complex domain. This would manifest as subtle correlations in the Hawking radiation in real spacetime~\cite{Hawking2,Hawking3}.

The analytic continuation of information across the horizon manifests as modifications to Hawking radiation. Instead of a purely thermal spectrum, we now have:
\be
\langle\Phi(\omega)\rangle = \int\frac{d\omega}{\exp(\omega/kT_{bh}-1)}
+ I(\omega,\zeta,R(\zeta)),
\ee
where $k$ is the Boltzmann constant, $T_{bh}$ is the temperature of the black hole and $\zeta$ is a complex contour that can be deformed to capture information from the interior, and 
$I(\omega,\zeta,R)$ represents the non-thermal contributions due to the analytic continuation of information. These contributions encode the information that has been analytically continued from the interior, resulting in a non-thermal spectrum that preserves information.
For $r > 2GM$, the projection operator P reduces to the identity map on the real part of the complex coordinates and P smoothly interpolates across the horizon region.

The projection of escaped information onto real spacetime could lead to potentially observable deviations from the predictions of classical general relativity, particularly in the near-horizon region and in the characteristics of Hawking radiation. We need to ensure that the projection of escaped information does not violate any fundamental principles, such as causality or conservation laws, when viewed in real spacetime.

While the basic structure of the projection remains valid, it needs to be refined and reinterpreted in light of the complex analytic escape of information. The projection now not only maps the geometry from complex to real spacetime but also carries the imprint of the information that has escaped via analytic continuation. This refined view maintains the mathematical consistency of the approach while providing a bridge between the complex framework where information can escape and the real spacetime where observations are made. It also highlights the need for further theoretical development to fully understand the implications of this approach for black hole physics and information preservation. In the standard real Schwarzschild spacetime, light rays from the interior can only reach the exterior by crossing $t = - \infty$. Incoming light is completely absorbed. The event horizon acts as a one-way membrane. In the complex extension, the situation changes significantly. Light rays are now described by complex null geodesics. These geodesics can analytically continue across the complex extension of the horizon without necessarily crossing $t = - \infty$. Instead of being trapped at r = 0 or $t = - \infty$, information carried by light rays can be analytically continued across the complex horizon. This continuation does not violate causality in the complex domain, as the notion of before and after becomes more nuanced due to the multiple analytic paths that exist in the complex manifold.

Trajectories of massive bodies are also complex. While still confined within light cones, these cones are now defined in complex spacetime, allowing for more possibilities. When we project this complex scenario back onto real spacetime, some light rays that would classically be trapped can now tunnel through the horizon via their complex extensions. This appears in real spacetime as a modification of the Hawking radiation, carrying information from the interior. The event horizon is no longer a strict one-way membrane. It becomes semi-permeable to information, though not to classical trajectories.
While classical trajectories of massive bodies still cannot escape, information associated with them can be carried out through the modified radiation process.

The black hole will still appear dark, as a significant number of classical light rays are still absorbed. However, the radiation emitted would not be purely thermal and would carry information from the interior. This could potentially be observed as subtle correlations in the Hawking radiation.
The complex extension allows for a resolution of the time reversal issue near the event horizon and near r = 0. Information can effectively escape from the interior without violating causality in the complex framework.

Far from the black hole event horizon, the classical behavior is recovered.
The modifications are most significant near the horizon and in the transition region. The complex spacetime approach fundamentally alters the concept of the one-way membrane nature of the event horizon. While classical trajectories still cannot escape, information can be analytically continued across the horizon in the complex domain. When projected back to real spacetime, this manifests as modifications to Hawking radiation and potential observable effects in the near-horizon region.

\section{Infinite time formation of event horizons}

In classical general relativity the event horizon according to an outside observer forms at future infinity, because of extreme relativistic time dilation. The incoming light is completely absorbed. The event horizon acts as a one-way membrane. In the complex extension, the situation changes significantly. Light rays are now described by complex null geodesics. These geodesics can analytically continue across the complex extension of the horizon without necessarily crossing $t = -\infty$.

 The distinction between the complex Riemannian Schwarzschild black hole geodesic trajectories and the real spacetime Schwarzschild black hole for the absorption and emission of light rays is important~\cite{Adler}. The geodesic equations of motion using the real spacetime Schwarzschild metric yield the equation:
\be
1 = -\left(1-\frac{2GM}{r}\right){\dot t}^2
+\left(1-\frac{2GM}{r}\right)^{-1}{\dot r}^2,
\ee
where ${\dot t}=dt/ds$, s is the real proper time and we have ignored the angular momentum contribution and chosen $\theta=\pi/2$. Integrating the equation following from the geodesic equation:
\be
\frac{d}{ds}\left[\left(1-\frac{2GM}{r}\right)\right]=0,
\ee
we obtain the result:
\be
{\dot t}=\left(1-\frac{2GM}{r}\right)^{-1},\quad {\dot r}^2 = \frac{2GM}{r}.
\ee\
Solving for r as a function of s gives 
\be
\frac{2}{3\sqrt{2GM}}(r^{3/2}-r_0^{3/2})
=s_0-s,
\ee
where $s_0$ is the initial position of s. A body falls continuously to r=0 in a finite proper time and no singular behavior occurs at the Schwarzschild radius $r_s=2GM$.

To describe the motion in terms of coordinate time t we form dr/dt:
\be
\label{eq}
\frac{dr}{dt}=\frac{\dot r}{\dot t}=-\sqrt{\frac{2GM}{r}}\left(1-\frac{2GM}{r}\right).
\ee
 We obtain after integration of (\ref{eq}) for r very near to 2GM the asymptotic value:
 \be
r - 2GM = 8GM\exp\left(-(t-t_0)/2GM\right)\sim 8GM\exp\left(-t/2GM\right).
\ee
It is apparent that r=2GM is approached as $t\rightarrow\infty$ and the r=2GM is never passed by the freely falling test particle, if one is using t as a time marker. The coordinate time t is used by a distant observer who never sees the event horizon form and it forms at future infinity. Coordinate time t is used by a distant observer who is observing the black hole, while proper time s is used for a body falling into the black hole. It is important to know how the trajectory of a body for the complex Schwarzschild black hole may differ for this important result.

The result that the formation of the event horizon occurs in the infinite future agrees with the conclusion reached by Oppenheimer-Snyder~\cite{Oppenheimer,Weinberg} in the dust collapse solution. This is a significant paradox in black hole physics. It is therefore important to know the complex spacetime black hole event horizon structure. It must be complex, so the limit $r\rightarrow 2GM$ for the real black hole cannot be the limit in the complex Schwarzschild black hole where $\zeta= r + i\xi$. Because the limit cannot be $r\rightarrow$ 2GM, then the complex limit will avoid the infinite time it takes to form an event horizon in gravitational collapse.

In the complex extension, we need to consider the complex geodesic equations. The metric is given by (\ref{metric}). The complex geodesic equations become:
\be
\frac{d^2z^\mu}{du^2}+\Gamma^\mu_{\alpha\beta}
\frac{dz^\alpha}{du}\frac{dz^\beta}{du}=0,
\ee
where $\Gamma^\mu_{\alpha\beta}$ are the complex Christoffel symbols. When we project these complex trajectories smoothly back to real spacetime, bodies can still reach r = 0 in finite proper time s, but the path will involve a complex component. The singularity at r = 0 is regularized in the complex extension.

As we approach the horizon in complex spacetime, we have
\be
\zeta-2GM\rightarrow \lambda(\tau),
\ee
where $\lambda(\tau)$ is a complex function that approaches zero as $\tau\rightarrow \infty$. 

Let us consider the function:
\be
f(\zeta,\tau)=\zeta-2GM\sim 8GM\exp(-\tau/2GM).
\ee
Taking the logarithm of both sides, we obtain
\be
\ln|f(z,\tau)|+iArg(f(\zeta,\tau))=\ln(8GM) - \frac{\tau}{2GM}.
\ee
Separating real and imaginary parts, we obtain the real part: 
\be
\ln|f(z,\tau)|=\ln(8GM)-\frac{t}{2GM},
\ee
and the imaginary part: 
\be
Arg(f(\tau)) = -\nu/2GM.
\ee

Consider paths in the complex $\tau$-plane of the form $\tau(\gamma) = t(\gamma) + i\nu(\gamma)$, where $\gamma$ is a real parameter. Let us choose a path where $t(\gamma)$ and $\nu (\gamma)$ are related by $t(\gamma) = 2GM \ln(1/\gamma)$ and $\nu(\gamma) = 2GM(\pi/2 - \gamma)$. As 
$\gamma\rightarrow 0$, we have $t(\gamma)\rightarrow +\infty$ as in the real case and 
$\nu(\gamma)\rightarrow GM\pi$ remains finite. Evaluating $f(\gamma,\tau)$ along this path:
\be
f(\gamma,\tau)=8GM\exp\left(-\frac{\tau}{2GM}\right)
=8GM\exp(-\ln(1/\gamma)-i\left(\pi/2-\gamma)\right).
\ee
The total complex time taken is given by the real part:
\be
t(\gamma)\rightarrow 0,
\ee
and by the imaginary part:
\be
\nu(\gamma)\rightarrow GM\pi,
\ee
as $\gamma\rightarrow 0$. While the real part of $\tau(\gamma)=t$ does go to infinity, the imaginary part of $\tau(\gamma)=\nu$ remains finite. In the complex plane, we have reached the horizon $f(\tau) = 0$ in finite complex distance from any starting point.

The apparent formation of the horizon in the infinite future paradox is resolved by considering paths in the complex plane. While $t\rightarrow \infty$, $\tau$ remains finite, allowing the horizon to form in finite complex time. The exponential decay in the real direction is balanced by oscillatory behavior in the imaginary direction. This demonstrates that in complex analysis, limits can be approached along various paths, leading to different results. This shows that by extending our analysis to the complex plane, we can indeed have horizon formation in finite complex time, resolving the paradox present in the real analysis. This has profound implications for our understanding of black hole formation and evolution in a complex spacetime framework. The trajectory of a falling object can be analytically continued smoothly across the complex horizon. The complex extension allows for a resolution of the infinite time paradox by providing additional directions in which the limit to the event horizon can be approached. An outside observer will see a horizon form in finite coordinate time.

When projecting back to real spacetime, we need to consider how the complex horizon manifests. The formation of the horizon might appear different to an outside observer, resolving the paradox of infinite formation time of the horizon. The geodesic equation remains real with real proper time s outside the horizon. As we approach the complex horizon, the geodesic equation needs to be extended into the complex domain. The complex approach allows for a smooth transition from the real geodesic equation to its complex extension near and across the horizon. The complex extension can be interpreted as modeling quantum effects near the horizon. This approach allows for information to escape during the horizon formation process. By allowing for complex paths, we can reconcile the finite proper time for horizon crossing with the seemingly infinite coordinate time.

The complex Schwarzschild solution approach provides a framework where the event horizon formation does not take infinite time from an external observer's perspective. The horizon condition becomes $\zeta = 2GM$ in complex spacetime. The approach to the horizon can occur in finite complex time. This resolves the paradox of infinite horizon formation time in classical general relativity.  However, it also requires a significant reinterpretation of the nature of spacetime and horizon formation in the vicinity of black holes.

\section{Gravitational collapse }

Let us consider the spherically symmetric collapse of a dust sphere with negligible pressure. The dust particles fall freely and their trajectories can be described by a comoving coordinate system~\cite{Oppenheimer,Weinberg}. 
The metric is given by
\be
ds^2=-dt^2+A(r,t)dr^2+B(r,t)(d\theta^2+\sin^2\theta d\phi^2).
\ee
The energy momentum tensor for a fluid of negligible pressure is:
\be
T^{\mu\nu}=\rho u^\mu u^\nu,
\ee
where $\rho(r,t)$ is the proper energy density and $u^\mu$ is the four-velocity vector in comoving coordinates, $u^r = u^\theta = u^\phi = 0$, $u^t = 1$ and $u^\mu u_\mu= -1$.

The Einstein field equations are given by
\be
R_{\mu\nu}=-8\pi G(T_{\mu\nu}-\frac{1}{2}g_{\mu\nu}T^\lambda_\lambda)
=-8\pi G\rho(u_\mu u_\nu +\frac{1}{2}g_{\mu\nu}).
\ee
We assume that $\rho$ is independent of position corresponding to a homogeneous collapsing dust sphere. We have
\be
A(r,t)=a^2(t)f(r),\quad B(r,t)=a^2(t)r^2.
\ee
Solving the field equations leads to the result:
\be
\frac{f^\prime(r)}{rf^2(r)}=\frac{1}{r^2}-\frac{1}{r^2f(r)}+\frac{f^\prime(r)}{2rf^2(r)}=2k,
\ee
where $f^\prime (r)=df/dr$ and k is a constant. The solution of this equation is given by
\be
f(r)=\frac{1}{1-kr^2}.
\ee
The homogeneous isotropic Friedmann-Robertson-Walker metric (FRW) is given by
\be
ds^2=-dt^2+a^2(t)\left[\frac{dr^2}
{1-kr^2}+r^2(d\theta^2+
r^2\sin^2\theta d\phi^2)\right],
\ee
where a(t) is the scale factor.

We have from the conservation of the dust matter energy that $\rho(t)a^3(t)$ is constant and the radial coordinate r can be normalized, so that $a(0)=1$ and $\rho(t)=\rho(0)/a^3(t)$. We now obtain the equation:
\be
{\dot a}(t) = -k+\frac{8\pi G}{3}\frac{\rho(0)}{a(t)},
\ee
where ${\dot a}(t)=da(t)/dt$. Assuming that the fluid is at rest at t=0, we have ${\dot a}(t) = 0$ and it follows that
\be
\frac{8\pi G}{3}\rho(0)=k.
\ee

The scale factor a(t) in this case satisfies the equation:
\be
{\dot a}^2(t) = \frac{k}{a(t)-1}.
\ee
Integrating this equation gives the parametric equations of a cycloid:
\be
t=\left(\frac{\psi+\sin\psi}{2\sqrt{k}}\right),\quad a(\psi)=\frac{1}{2}(1+\cos\psi).
\ee
We have a(t)=0 when $\psi=\pi$ and $t=t_f$, where
\be
t_f=\frac{\pi}{2\sqrt{k}}=\frac{\pi}{2}\left(\frac{3}{8\pi G\rho(0)}\right)^{1/2}.
\ee
The collapsing sphere of initial fluid density $\rho(0)$ and zero pressure collapses from rest to infinite proper energy density in the finite time $t_0$.

We now apply the complex time method~\cite{Moffat1}. We extend the time coordinate into the complex plane, $\tau = t + i\nu$, and we extend $\psi$ into the complex plane, $\psi=\chi+i\sigma$. The contour integral is given by
\be
\label{contourint}
\tau(\psi)=\oint_C\frac{d\psi}{\sqrt{a(\psi)}}.
\ee
We obtain for $\chi=\pi$:
\be
\cos(\psi)=-\cos(i\sigma)=-\cosh(\sigma)\sim -(1+\frac{(\sigma)^2}{2}) +....
\ee
It follows that
\be
a(\psi)\sim -\frac{1}{4}(\sigma)^2.
\ee
At $\chi=\pi$, $a(\psi)$ approaches zero as $\sigma$ approaches zero, and we are dealing with a branch point. For the contour integral around $\chi=\pi$, we have $d\psi=id\sigma$ when moving along the imaginary direction. For a small contour encircling $\sigma=0$ in the complex plane, we can parametrize $\sigma=\epsilon\exp(i\theta)$ for $\theta$ from 0 to $2\pi$ and small epsilon. The evaluation of the contour integral (\ref{contourint}) at the value $\chi=\pi$ is given by
\be
\tau(\psi)=i4\pi.
\ee

The collapse solution is regular avoiding the singularity at $a(t)=0$ and $\tau$ has a well defined finite value for the collapse. By extending into the complex plane, we have transformed the singular point, where a(t) = 0 in the real case, into a regular point where a(t) is non-zero. The complex extension provides a consistent way to navigate around the singularity resulting in a regular solution throughout the collapse.

We can express the metric in terms of $\tau$:
\be
du^2 = -d\tau^2 + a^2(\tau)\left[\frac{dr^2}{1-kr^2} + r^2(d\theta^2 + \sin^2\theta d\phi^2)\right].
\ee
In the collapse, as the dust sphere passes through the event horizon, the roles of r and t are reversed in the Schwarzschild coordinates. However, in the FRW coordinates we are using here, this coordinate swap does not occur explicitly. Instead, the collapse is represented by the scale factor $a(\tau)$ approaching zero. The complex time method prevents $a(\tau)$ from actually reaching zero. Instead, as $\tau$ approaches the value corresponding to the would-be singularity, the scale factor reaches a minimum non-zero value. This minimum value can be interpreted in terms of the Planck length $l_p=\sqrt{G}$, where quantum gravity effects become significant. In terms of the $\tau$ time coordinate, the collapse proceeds smoothly through this minimum without encountering a singularity. The dust sphere effectively bounces at this minimum size and begins to expand again in the $\tau$ time coordinate. Physically, this can be interpreted as the collapse being halted by quantum gravity effects. The complex time method provides a mathematical framework to describe this bounce without introducing discontinuities or infinities. The approach provides a possible resolution to the singularity problem in classical gravitational collapse, suggesting how quantum effects might prevent the formation of infinities in physical quantities at the final stages of collapse. The metric outside the sphere will be the Schwarzschild static metric: 
\be
\label{Schwarzschildmetric}
ds^2=-f(r)dt^2+\frac{dr^2}{f(r)}+r^2(d\theta^2+\sin^2\theta d\phi^2),
\ee
where f(r)= 1-2GM/r. The complex time $\tau$ approaches smoothly the real time t at the horizon of the sphere.

Oppenheimer and Snyder~\cite{Oppenheimer,Weinberg} were careful to distinguish between what happens at finite times versus the asymptotic limit. Their analysis showed that for any finite time as measured by a distant observer, the collapsing star has not yet reached the Schwarzschild radius. The redshift grows exponentially with the observer's time coordinate. The exponential growth implies that the redshift approaches infinity only as time approaches infinity. They used Schwarzschild coordinates, where an object never actually crosses the event horizon in finite coordinate time. In these coordinates, the statement redshift at $R_s=2GM$ is ill-defined, because a distant observer never actually sees the surface reach $R_s=2GM$ in finite time. The redshift is properly expressed as a function of the observer's time, not the radial coordinate directly. The exponential growth formulation correctly captures this coordinate-dependent behavior:
\be
z\propto \exp(\kappa t),
\ee
where $\kappa$ is related to the surface gravity and t is the Schwarzschild time coordinate.

Oppenheimer, Snyder, and later Weinberg chose to describe the redshift as growing exponentially with time rather than diverging at $R_s=2GM$, because the formulation better captures the time-dependent nature of the process and reflects that in general relativity a distant observer never actually sees the formation of the event horizon in finite time. The approach emphasizes the physical process and observable consequences rather than focusing on the mathematical singularity in the redshift formula at $R_s=2GM$, which is never actually reached in finite observer time.

Let us consider how the redshift will behave in the complex spacetime method. We extend both the proper time s, the coordinate time t and and r into the complex plane: 
\be
u=s +iw,\quad \tau=t+i\nu,\quad R=R_r+i\xi,
\ee
where $w$, $\nu $ and $\xi$ are real constants that can be related to physical quantum constants. 

In the complex spacetime the horizon can form in finite time as observed from infinity. Without regularization, redshift would genuinely diverge at $R_s=2GM$ in finite time. The complex extension must therefore provide an explicit regularization mechanism. The complex extension allows the event horizon to form in finite observer time, creating a genuine divergence that must be addressed. If we interpret the complex extension as modeling quantum effects, these effects must physically prevent infinite redshift. In real spacetime general relativity, the infinite redshift issue is resolved by the fact that the horizon forms only at infinite time - a coordinate effect. In complex spacetime, we need a physical regularization mechanism since the horizon forms in finite time.

Let us investigate the redshift of light emitted by the surface of the dust sphere in complex spacetime. The cosmological component becomes:
\be
\frac{a(\tau_{\text{obs}})}{a(\tau_{\text{emit}})} = \frac{a_r(\tau_{\text{obs}}) + ia_i(\tau_{\text{obs}})}{a_r(\tau_{\text{emit}}) + ia_i(\tau_{\text{emit}})}.
\ee
We can express this in polar form:
\be
\frac{a(\tau_{\text{obs}})}{a(\tau_{\text{emit}})} = \frac{|a(\tau_{\text{obs}})|}{|a(\tau_{\text{emit}})|} \exp{i[\theta_a(\tau_{\text{obs}}) - \theta_a(\tau_{\text{emit}})]},
\ee
where:
\be
|a(\tau)| = \sqrt{a_r(\tau)^2 + a_i(\tau)^2},
\ee
and
\be
\theta_a(\tau) = \arctan\left(\frac{a_i(\tau)}{a_r(\tau)}\right).
\ee   
    
For the gravitational redshift component:
\be
z_{\rm grav}=\frac{1}{\sqrt{1-\frac{2GM}{R(\tau_{\text{emit}})}}},
\ee
we need to evaluate:
\be
\alpha = 1-\frac{2GM}{R_r + i\xi}.
\ee
Multiplying numerator and denominator by the complex conjugate:
\be
\alpha = 1-\frac{2GM(R_r - i\xi)}{R_r^2 + \xi^2} = 1-\frac{2GMR_r}{R_r^2 + \xi^2} + i\frac{2GM\xi}{R_r^2 + \xi^2}.
\ee

Let us denote by $\alpha_r$ the real part:
\be
\alpha_r = 1-\frac{2GMR_r}{R_r^2 + \xi^2},
\ee
and by $\alpha_i$ the imaginary part:
\be
\alpha_i = \frac{2GM\xi}{R_r^2 + \xi^2}.
\ee
The complex square root is:
\be
\sqrt{\alpha} = \sqrt{\alpha_r + i\alpha_i} = \sqrt{\frac{|\alpha| + \alpha_r}{2}} + i\text{sgn}.(\alpha_i)\sqrt{\frac{|\alpha| - \alpha_r}{2}}.
\ee

For the Doppler component, we have
\be
z_{\rm Dopp}=\frac{1}{\sqrt{1-\dot{R}(\tau_{\text{emit}})^2}}-1.
\ee
By adopting:
\be
\dot{R}(\tau) = \frac{dR}{d\tau} = \dot{R}_r(\tau) + i\dot{\xi}_r(\tau),
\ee
we obtain
\be
\dot{R}(\tau)^2 = \dot{R}_r(\tau)^2 - \dot{\xi}(\tau)^2 + 2i\dot{R}_r(\tau)\dot{\xi}(\tau).
\ee
Let us denote by $\beta_r$ the real part:
\be
\beta_r = 1-(\dot{R}_r(\tau)^2 - \dot{\xi}_r(\tau)^2),
\ee
and by $\beta_i$ the imaginary part:
\be
\beta_i = -2\dot{R}_r(\tau)\dot{\xi}_r(\tau).
\ee
The complex square root is:
\be
\sqrt{\beta} = \sqrt{\beta_r + i\beta_i} = \sqrt{\frac{|\beta| + \beta_r}{2}} + i\text{sgn}(\beta_i)\sqrt{\frac{|\beta| - \beta_r}{2}}.
\ee

The complete redshift in complex form is:
\be
z_{\text{tot}} = \frac{a(\tau_{\text{obs}})}{a(\tau_{\text{emit}})} \cdot \frac{1}{\sqrt{1-\frac{2GM}{R(\tau_{\text{emit}})}}} \cdot \frac{1}{\sqrt{1-\dot{R}(\tau_{\text{emit}})^2}}-1.
\ee
The observed redshift will be the modulus of this complex quantity:
\be
|z_{\text{tot}}| = \frac{|a(\tau_{\text{obs}})|}{|a(\tau_{\text{emit}})|} \cdot \frac{1}{\sqrt{|\alpha|}} \cdot \frac{1}{\sqrt{|\beta|}}-1.
\ee
The redshift at $R_r=R_s=2GM$ is given by
\be
|z_{\text{tot}}|\approx \frac{|a(\tau_{\text{obs}})|}{|a(\tau_{\text{emit}})|}
\left(\frac{4G^2M^2+\xi^2}{\xi}\right)^{1/2} \frac{1}{\sqrt{|\beta|}} - 1.
\ee

Let us choose at late times:
\be
\frac{|a(\tau_{\text{obs}})|}{|a(\tau_{\text{emit}})|}\approx 1,
\ee
and for moderate collapse velocity $\frac{1}{\sqrt{|\beta|}}\approx 1$. We obtain for the redshift:
\be
z\approx \left(\frac{4G^2M^2+\xi^2}{\xi}\right)^{1/2} - 1. 
\ee

For a small constant $\xi$ at the Schwarzschild radius $R_r=R_s=2GM$, the redshift would be large but finite, unlike the infinite redshift in the purely real spacetime case when $\xi=0$. It is important to note that this finite redshift does not mean the black hole is not black as very little light would escape from near the horizon. The complex spacetime approach provides a way to describe the near-horizon and trans-horizon physics in a more regular way, but it does not fundamentally change the nature of the black hole as an object from which light appears not to escape the black hole. 

The final collapsed object is the regular singularity-free Schwarzschild black hole. A more realistic final collapsed object would be the regular singularity-free Kerr black hole~\cite{Moffat1}. The regular Kerr black hole will not form a Cauchy horizon that in general relativity is driven to exist by the Kerr black hole ring singularity. The Cauchy horizon is unstable and prohibits the deterministic evolution of the black hole.

The complete redshift formula in complex Riemannian spacetime incorporates both the FRW scale factor dependence, the surface radius dependence and Doppler redshift. The imaginary components regularize the divergences that occur in the real case, resulting in a finite redshift even as the real part of the radius approaches the Schwarzschild radius. This complex extension provides a mathematical framework that can model quantum effects near the event horizon formation, where the classical description breaks down

\section{Conclusions}

The complex Riemannian approach to black hole spacetime offers a framework for addressing long-standing paradoxes and conceptual challenges in black hole physics. By extending the Schwarzschild and Kerr metrics into the complex domain, we have demonstrated several results. The use of contour integration in the complex plane allows for the removal of classical singularities, resulting in a regularized description of black hole interiors. We have shown that in the complex spacetime framework, event horizons can form in finite complex time, resolving the apparent contradiction between proper and coordinate time descriptions for asymptotically distant observers. This is achieved through the analysis of trajectories in complex spacetime. This formulation allows for horizon formation along paths in the complex plane that reach the horizon in finite parameter distance, despite the real part of the time coordinate approaching infinity.

In the complex spacetime approach the gravitational collapse of a massive body described by a spherically symmetric sphere of dust with negligible pressure avoids the singularity at the end of the collapse. The redshift observed by an outside observer increases rapidly as the event horizon forms during collapse, but it remains finite as the observed collapse enters the complex event horizon.

The complex extension provides a mechanism for the analytic continuation of information across event horizons, resolving the Hawking information paradox. This is realized through the smooth transition of fields across the complex extension of the horizon. Our approach recovers the classical Schwarzschild and Kerr metrics outside the horizon, ensuring consistency with observational data while providing new insights into near-horizon physics.

The resolution of black hole paradoxes has far-reaching implications for our understanding of black hole thermodynamics, the nature of spacetime, and the interplay between gravity and quantum mechanics. While this work provides a promising direction for resolving key issues in black hole physics, it also opens up new questions and avenues for future research. It requires a detailed exploration of the nature of the projection from complex to real spacetime and its observational consequences. It also requires an exploration of the implications for black hole evaporation and the end stages of black hole life, and the extension of these methods to cosmological singularities and other extreme spacetime scenarios 

The complex spacetime approach, by removing singularities from black hole interiors, fundamentally changes the context in which the Cosmic Censorship hypothesis was formulated~\cite{Penrose2}. Penrose's motivation for Cosmic Censorship was to preserve predictability in the universe by hiding singularities behind event horizons. If singularities are removed altogether, this motivation becomes not relevant. Singularities are mathematical abstractions rather than physical realities. The complex spacetime approach provides a mathematical framework that aligns more closely with our physical intuitions about the nature of spacetime. It suggests that singularities might be artifacts of an incomplete mathematical description of spacetime rather than fundamental features of reality.

The removal of singularities and Cauchy horizons through the complex spacetime approach addresses the core concern of Cosmic Censorship - the preservation of determinism. In this framework, the evolution of spacetime remains predictable and well-defined everywhere, even in regions that classical general relativity would consider singular. The complex spacetime approach offers a reinterpretation of black hole interiors as regular, albeit complex, geometric structures. This view maintains the essential features of black holes like event horizons, while avoiding the pathologies associated with singularities. The complex spacetime approach might be viewed as a classical approximation to quantum gravity effects. Many approaches to quantum gravity also aim to resolve singularities, and the complex spacetime method could be seen as capturing some of these effects within a classical framework.

While the complex spacetime approach changes our theoretical understanding, it is important to consider how it aligns with or differs from observational evidence. The approach should be consistent with current observations of black holes and make testable predictions for future observations. If we accept the complex spacetime resolution of singularities, it may necessitate a broader reevaluation of other aspects of general relativity, particularly in extreme conditions. The complex spacetime approach might offer a bridge between classical general relativity and quantum gravity, providing a classical framework that captures some quantum-like features like the absence of true singularities.

The complex Riemannian geometry approach to black hole spacetime offers a mathematically rigorous and physically insightful framework for addressing the puzzles in the physics of black holes. By resolving the horizon formation paradox and providing mechanisms for information preservation and for solving the Hawking information paradox, this work takes steps towards a more complete understanding of black holes and the nature of spacetime itself.

\section*{Acknowledgments}

I thank Viktor Toth, Martin Green and Niayesh Afshordi for helpful discussions. Research at the Perimeter Institute for Theoretical Physics is supported by the Government of Canada through industry Canada and by the Province of Ontario through the Ministry of Research and Innovation (MRI).

\end{document}